# RAG-based Architectures for Drug Side Effect Retrieval in LLMs

Shad Nygren[1], Pinar Avci[1], Andre Daniels[1], Reza Rassool[1], Afshin Beheshti[1,2,3], Diego Galeano[1,4]

[1]Kwaai, CA, USA.
[2]Center for Space Biomedicine, McGowan Institute for Regenerative Medicine, Department of Surgery, University of Pittsburgh, Pittsburgh, Pennsylvania, USA.
[3]Broad Institute of MIT and Harvard, Cambridge, Massachusetts, USA.
[4]Department of Electronics and Mechatronics Engineering, Facultad de Ingeniería, Universidad Nacional de Asunción - FIUNA, Luque, Paraguay.

**Abstract**

Drug side effects are a major global health concern, necessitating advanced methods for their accurate detection and analysis. While Large Language Models (LLMs) offer promising conversational interfaces, their inherent limitations—including reliance on black-box training data, susceptibility to hallucinations, and lack of domain-specific knowledge—hinder their reliability in specialized fields like pharmacovigilance. To address this gap, we propose two Retrieval Augmented Generation (RAG) and GraphRAG architectures, integrating comprehensive drug side effect knowledge into a Llama 3-8B Language Model. Through extensive evaluations on 19,520 drug-side effect associations (covering 976 drugs and 3,851 side effect terms), our results demonstrate that GraphRAG achieves near-perfect accuracy in drug side effect retrieval. This framework offers a highly accurate and scalable solution, signifying a significant advancement in leveraging LLMs for critical pharmacovigilance applications.

**Introduction**

Drug side effects represent a critical global public health challenge, significantly contributing to morbidity and mortality worldwide[1–4]. The rapid pace of drug development often outstrips the capacity of healthcare professionals to stay abreast of new medication side effects, particularly outside their primary specialties[5,6]. This issue is further complicated when patients report potential adverse reactions, requiring physicians to swiftly assess causality during time-constrained appointments. Current tools, such as drug handbooks[7], electronic medical records (EMRs)[8], and Spontaneous Reporting Systems (FAERS)[9,10], while valuable, often prove time-consuming and limited in their search capabilities, underscoring the urgent need for more efficient and accessible resources for drug side effect assessment in clinical practice[5,11].

Large language models (LLMs)[12–17] offer a promising avenue with their intuitive, conversational interfaces, holding the potential to streamline clinical workflows and enhance decision-making. These models support semantic search and complex queries, enabling the identification of drugs or diseases associated with specific symptoms[18], which can provide insights into potential causal relationships. Despite these advancements in natural language processing, the application of off-the-shelf LLMs in domain-specific tasks such as drug side effect identification has yielded mixed results[19–21], frequently struggling with accuracy and reliability in specialized fields like pharmacovigilance[14,21]. Their limitations stem from knowledge constrained by black-box training data, a propensity for hallucinations, and a general lack of domain-specific expertise, hindering their effectiveness in handling nuanced medical data and generating contextually appropriate insights.

To overcome these significant challenges, we propose two novel architectures designed to integrate domain knowledge about drug side effects into a Llama 3-8B Language Model: Retrieval Augmented Generation (RAG) and GraphRAG. Our first architecture employs RAG, which enhances LLMs by retrieving relevant information from an external Pinecone vector database where drug side effect information is stored as feature vectors. The second architecture utilizes GraphRAG, which leverages a Neo4j graph database to store and efficiently handle more complex relationships of drug side effect associations. Both frameworks incorporate custom split functions and filtering modules to optimize user prompts for accurate retrieval. Through extensive evaluations on 19,520 associations between 976 marketed drugs and 3,851 unique side effect terms, we demonstrate that GraphRAG achieves near-perfect accuracy in drug side effect retrieval, significantly outperforming standalone LLMs and standard RAG approaches. This underscores the transformative potential of our framework in providing highly accurate information on drug side effects within the context of Large Language Models.

**Results**

**A Retrieval-Augmented Generation (RAG) framework for drug side effect retrieval**

Our Retrieval Augmented Generation (RAG) system was engineered for efficient drug side effect retrieval, leveraging the Side Effect Resource (SIDER) 4.1 database as its primary knowledge source. This database, compiled from adverse event reports in randomized controlled trials and post-marketing surveillance, was initially filtered to include drugs with known Anatomical, Therapeutic, and Chemical (ATC) classifications and side effects categorized as MedDRA Preferred Terms (PT). This yielded a dataset of 141,209 associations, linking 1,106 marketed drugs to 4,073 unique side effect terms (**Fig. 1a**). However, due to the extensive number of these associations, utilizing the full dataset for comprehensive evaluation would have been computationally prohibitive. Therefore, for our assessment, we subsequently created a balanced subset of 19,520 drug-side effect pairs, as detailed in the Evaluation Dataset Creation section of the **Methods**.

To facilitate text-based retrieval, the raw SIDER data was processed into two distinct text formats (**Fig. 1b**). "Data Format A" provides a structured, comma-separated list of all known side effects for a given drug (e.g., "The drug aspirin causes the following side effects or adverse reactions: shock, peptic ulcer, contusion, …"). In contrast, "Data Format B" presents each drug-side effect pair on a new line, enhancing granularity (e.g., "The drug aspirin may cause urticaria as an adverse effect, adverse reaction, or side effect.").

For the RAG pipeline (**Fig. 1c, d**), "Data Format A" was segmented into chunks using a custom algorithm that splits text at newlines. These chunks were then embedded into a 1,536-dimensional vector space utilizing the OpenAI ada002 embedding model, chosen for its capacity to support up to 8,192 tokens, sufficient for even the longest text chunks in Format A (exceeding 10,000 characters). The resulting embeddings were indexed in a Pinecone vector database, enabling rapid similarity-based retrieval.

The RAG query workflow operates as follows: an end-user query (e.g., "Is urticaria an adverse effect of aspirin?") is first embedded using the ada002 model and compared against the top five most similar entries in the Pinecone database. Concurrently, an entity recognition module extracts drug and side effect terms (e.g., "aspirin" and "urticaria") from the query prompt. A subsequent filtering module checks if the identified drug-side effect pair from the query exists within the top five retrieved results. Based on this check, a modified prompt is generated: if a match is found, the prompt states that the drug is known to be associated with the side effect; otherwise, it specifies that the drug is not known to be associated with the side effect. Our modified prompt has the following structure:

```
"You are asked to answer the following question with a single word: YES or
NO. Base your answer strictly on the RAG Results provided below. After your
YES or NO answer, briefly explain your reasoning using the information from
the RAG Results. Do not infer or speculate beyond the information provided.
Question:\n\n" + query + rag_results
```

The variable `rag_results` contains the result from RAG. For instance, it can be:

```
"No, the side effect " + side_effect_query + " is not listed as an adverse
effect, adverse reaction or side effect of the drug " + drug_query
```

Where the drug and side effect query are the terms extracted using the entity recognition module.

The modified prompt is then passed to a language model (Llama 3, 8B parameters) which generates a binary YES/NO response. This binary output was specifically chosen because our evaluation framed drug side effect identification as a binary classification problem to predict the presence or absence of an association. The entire pipeline is orchestrated via AWS Lambda and API Gateway on Amazon Bedrock, ensuring scalability and seamless integration. This framework effectively addresses the binary classification task of distinguishing known from unknown drug-side effect associations

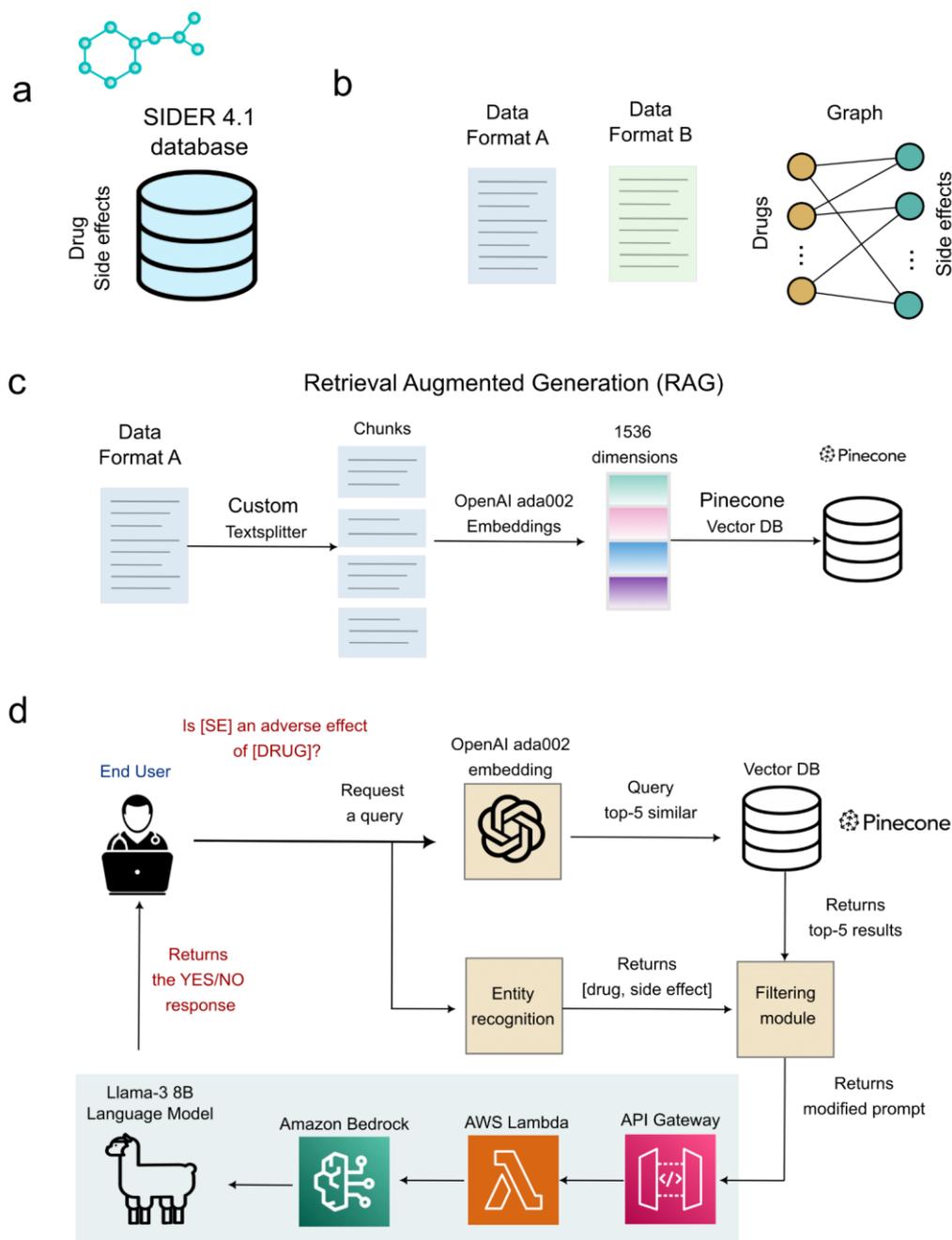

**Figure 1. Evaluation of drug side effect identification with Llama-3 8B-based models. (a)** 19,520 drug-side effect associations (covering 976 drugs and 3,851 side effect terms) were extracted from the SIDER 4.1 database; **(b)** Drug side effect associations were stored in three formats: format A, format B, and as a relational graph; **(c)** Retrieval Augmented Generation (RAG) to extract drug side effect information from data format A and B; **(d)** Evaluation procedure. The user writes prompt queries, and we run three different language models (RAG on data A, RAG on data B, and graph RAG) to extract drug side effect associations. Each model is assessed based on its binary response.

**Graph-Based Retrieval Augmented Generation (GraphRAG) for Drug Side Effect Data**

In our GraphRAG framework, drug-side effect associations are precisely modeled as a graph-based representation, leveraging the extensive Side Effect Resource (SIDER) 4.1 database previously described. Within this structure (Fig. 2a), drugs and side effects constitute distinct nodes, and their known relationships are encoded as directed edges, specifically labeled `"may_cause_side_effect"`. This graph is implemented within a Neo4j database, a robust graph management system that facilitates efficient querying via Cypher, enabling rapid traversal and retrieval of complex drug-side effect relationships.

The GraphRAG system is designed to process user queries, such as "Is headache an adverse effect of metformin?" (**Fig. 2b**). The workflow begins with an entity recognition module extracting drug and side effect terms (e.g., "metformin" and "headache") from the submitted query. These extracted entities are then used to construct a precise Cypher query, showcase with an example below:

```
cypher = f"""
    MATCH (s)-[r:May_Cause_Side_Effect]->(t)
    WHERE s.name = 'metformin' AND t.name = 'headache'
    RETURN s, r, t
    """
```

which is executed against the Neo4j database. This query efficiently identifies the presence or absence of a direct edge between the specified drug and side effect nodes, returning matching associations or an empty result accordingly.

The retrieved results are then processed by a prompt engineering module to generate a context-specific input for the language model. If a match is found, the prompt is modified to state, "Metformin is known to be associated with headache as a side effect". Conversely, if no association is found, the prompt states, "Metformin is not known to be associated with headache as a side effect". This prompt modification strategy is identical to that employed in our RAG architecture, ensuring a consistent approach to informing the language model. This refined prompt is fed into a Llama-3 (8B parameters) model, which generates a binary YES/NO response. As with our RAG framework, this binary output is a deliberate choice, aligning with our evaluation's formulation of drug side effect identification as a binary classification problem to predict the presence or absence of an association. The entire system is orchestrated using AWS Lambda and API Gateway on Amazon Bedrock, ensuring scalability and seamless integration for real-time query handling.

This GraphRAG approach offers distinct advantages over traditional text-based retrieval methods for pharmacovigilance. By representing drug-side effect associations as a graph, it enables highly precise, relationship-driven queries that significantly reduce ambiguity and enhance retrieval accuracy. Its integration with Neo4j facilitates complex traversals, such as identifying indirect associations or clustering related side effects, while the use of Llama-3 ensures clear, user-friendly responses. Deployed on robust AWS infrastructure, this framework effectively combines the strengths of graph-based data structures and advanced language modeling, providing a scalable and highly accurate solution for drug side effect retrieval in both clinical and research settings.

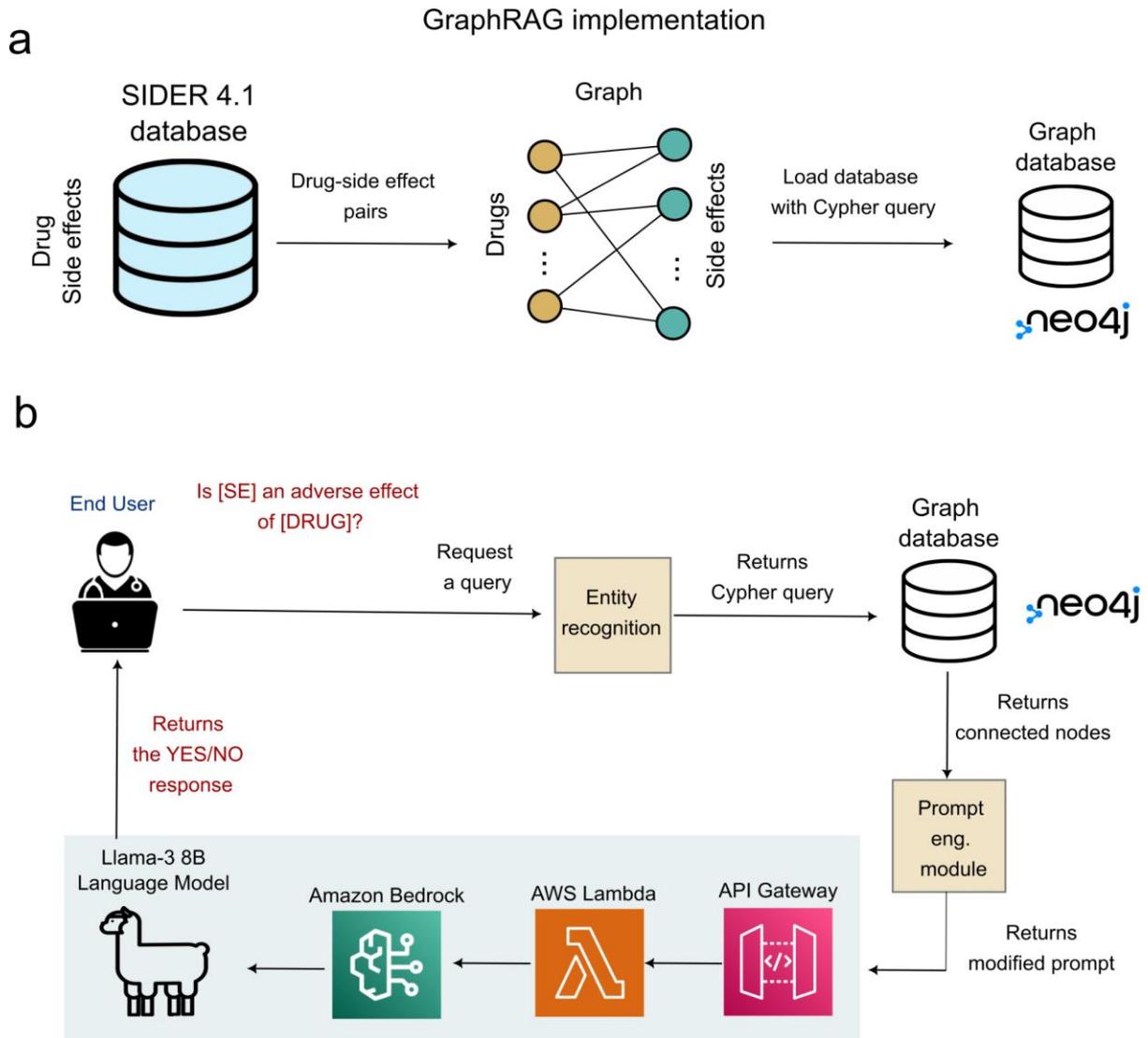

**Figure 2 - GraphRAG framework for drug side effect retrieval.** (a) The SIDER 4.1 database, containing drug-side effect pairs, is transformed into a graph structure where drugs (orange nodes) and side effects (green nodes) are connected by "may_cause_side_effect" edges, and loaded into a Neo4j graph database using Cypher queries. (b) Workflow of the GraphRAG system: an end-user submits a query (e.g., "Is [SE] an adverse effect of [DRUG]?"), which undergoes entity recognition to extract drug and side effect terms. A Cypher query retrieves matching associations from the Neo4j database, and a prompt engineering module refines the input for the Llama-3 8B language model, hosted on Amazon Bedrock with AWS Lambda and API Gateway, to generate a binary YES/NO response.

**Performance Evaluation of RAG and GraphRAG Frameworks**

To quantitatively assess the effectiveness of our LLM-based system in retrieving drug-side effect associations, we designed a rigorous binary classification task aimed at predicting the presence or absence of such associations (YES/NO response). This evaluation framework in in line with the drug side effect prediction literature[22,23]. For this evaluation, we prepared a balanced dataset from the Side Effect Resource (SIDER) 4.1 database. We focused on drugs with at least ten known side effect associations to ensure robust analysis. For each selected drug, we constructed a positive set by randomly sampling ten known side effects and an equally sized negative set by sampling side effects not associated with that drug. This meticulous process yielded a balanced dataset comprising 19,520 drug-side effect pairs, encompassing 976 marketed drugs and 3,851 unique side effect terms, all categorized as MedDRA Preferred Terms.

We rigorously assessed the performance of our GraphRAG and RAG frameworks against a standalone Llama 3-8B model across standard binary classification metrics: accuracy, F1 score, precision, sensitivity, and specificity. As depicted in Fig. 3, GraphRAG achieved near-perfect results, demonstrating an accuracy of 0.9999, F1 score of 0.9999, precision of 0.9998, sensitivity of 0.9999, and specificity of 0.9998. In stark contrast, the standalone Llama 3-8B model exhibited poor performance, with an accuracy of 0.529, an F1 score of 0.164, and a sensitivity of 0.092, underscoring its inherent limitations in accurately predicting drug-side effect associations without domain-specific knowledge augmentation. To further investigate whether this poor performance extends to significantly larger models, we also evaluated ChatGPT 3.5 and ChatGPT 4 on a subset of 51 randomly selected drugs. We observed a mean accuracy of approximately 0.55 for ChatGPT 3.5 and 0.63 for ChatGPT 4. This demonstrates that even advanced, larger language models struggle to accurately identify drug side effects for marketed drugs without specialized augmentation.

The RAG models also demonstrated marked improvements over the standalone Llama 3-8B. Notably, RAG with Data Format A achieved an accuracy of 0.886 and sensitivity of 0.776, while RAG with Data Format B performed significantly better, with an accuracy of 0.998 and sensitivity of 0.999. This disparity highlights the critical impact of data representation: Data Format B, which structures drug-side effect pairs individually, enables more precise retrieval compared to Data Format A, where side effects are aggregated into a single list per drug. These results collectively underscore the high effectiveness of both RAG and GraphRAG in significantly enhancing prediction accuracy and recall, especially when leveraging structured knowledge representations tailored to the task.

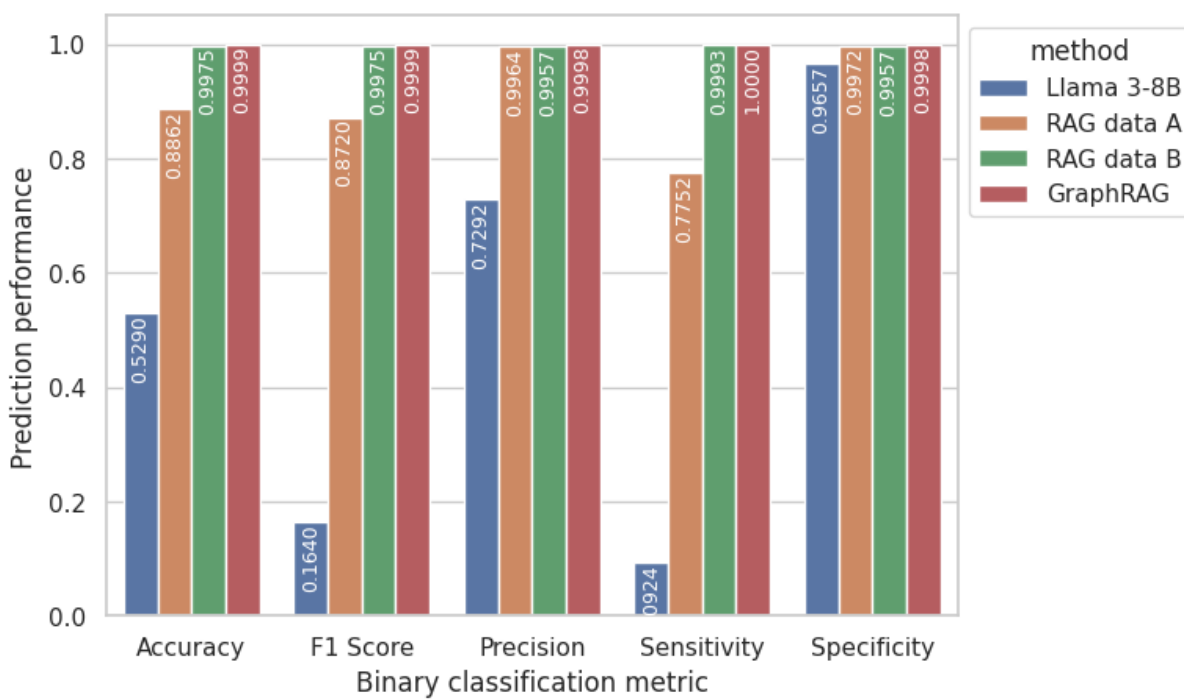

**Figure 3 – Binary classification metrics at drug-side effect association retrieval by different methods:** standalone Llama 3-8B, RAG data format A, RAG data format B and GraphRAG. The performance was assessed using a balanced dataset of 19,520 drug-side effect pairs, covering 976 marketed drugs and 3,851 side effect terms categorized as MedDRA Preferred Terms obtained from SIDER 4.0.

Further detailed analysis of prediction performance across drug categories, defined by the Anatomical Therapeutic Chemical (ATC) classification system, is provided in **Supplementary Figure 1**, focusing on accuracy. GraphRAG consistently outperformed all other models across every ATC category, showing particularly notable improvements in precision-critical domains such as antineoplastic agents and cardiovascular drugs. In these complex and diverse categories, GraphRAG maintained high accuracy and sensitivity without sacrificing specificity, robustly handling intricate drug-side effect relationships. Consistent near-perfect accuracy was also observed across various side effect MedDRA Organ classes, as shown in **Supplementary Figure 2**. These findings collectively emphasize GraphRAG's superior ability to integrate graph-based retrieval with advanced prompt engineering, effectively leveraging structured knowledge within a Neo4j graph database to achieve unparalleled prediction quality. By comprehensively modeling drug-side effect associations as a graph, GraphRAG successfully overcomes the limitations of generative language models in reasoning about complex biomedical relationships, setting a new benchmark for drug safety and pharmacovigilance applications by offering a scalable and highly accurate solution for identifying adverse drug effects in clinical and research settings.

**Code availability**

The data and code used in our study is available here: https://github.com/diegogalpy/RAG-based-models-for-drug-side-effect-retrieval

**Acknowledgments**

This project was supported by Virtual Hipster Inc. and Kwaai.

**Author Contribution**

D.G. conceived the study and supervised the project. S.N. and D.G. designed the RAG frameworks. S.N. implemented software architecture and ran experiments. P.A., A.D., and A.F. provided critical review and guidance on the medical and clinical aspects of the study. R.R. provided methods for analysis. All authors reviewed, edited, and approved of the final manuscript.

## Discussion

The pervasive challenge of drug side effects, which significantly contribute to global morbidity and mortality[1,2], necessitates highly accurate detection and analysis methods[24]. Despite the emergence of large language models (LLMs) as powerful tools, their application in specialized fields like pharmacovigilance has been hampered by limitations such as dependence on black-box training data, hallucination tendencies, and insufficient domain-specific knowledge. Traditional pharmacovigilance methods, though foundational, often struggle with data silos and fragmentation, further complicating comprehensive analysis. To bridge this critical gap, we developed and evaluated Retrieval-Augmented Generation (RAG) and GraphRAG frameworks. Our findings underscore the transformative potential of these AI-driven approaches, particularly GraphRAG, in achieving near-perfect accuracy for drug side effect retrieval by augmenting LLMs with structured domain knowledge.

Our study directly addresses these challenges by introducing Retrieval-Augmented Generation (RAG) and GraphRAG frameworks, leveraging the comprehensive Side Effect Resource (SIDER) 4.1 database[25]. By integrating structured knowledge with advanced LLMs, our systems achieve near-perfect performance in retrieving drug-side effect associations. Specifically, GraphRAG demonstrated an accuracy of 0.9999 and sensitivity of 0.9999 across 19,520 drug-side effect pairs. These findings underscore the transformative potential of AI-driven approaches in pharmacovigilance, enabling the precise and scalable retrieval of adverse event data.

A key limitation of conventional LLMs, exemplified by the poor performance of the standalone Llama 3-8B model and even larger models like ChatGPT 3.5 and 4, is their inability to effectively reason about complex biomedical relationships without domain-specific augmentation. While fine-tuning LLMs can offer some performance improvements[26], it is typically resource-intensive and often less accurate in nuanced medical contexts compared to retrieval-augmented methods. In contrast, our RAG and GraphRAG frameworks effectively overcome this by incorporating structured knowledge representations—text-based embeddings for RAG and sophisticated graph-based structures for GraphRAG. This integration resulted in significant performance gains (RAG Data Format B accuracy: 0.998; GraphRAG accuracy: 0.9999). GraphRAG, in particular, excelled by modeling drug-side effect associations as a bipartite graph within a Neo4j database, facilitating highly precise, relationship-driven queries. This capability proved invaluable in precision-critical domains such as antineoplastic agents and cardiovascular drugs, where GraphRAG consistently maintained high sensitivity and specificity across Anatomical Therapeutic Chemical (ATC) categories. These results firmly underscore the critical importance of structured knowledge in enhancing LLM reliability for medical applications, thereby establishing a new benchmark for drug safety monitoring.

Despite these significant advancements, our current framework has several limitations. The SIDER 4.1 database, while comprehensive for known associations, primarily captures reported side effects, which means many unreported or emerging adverse events, particularly those identified in the post-marketing phase, remain undetected. Underreporting is a pervasive issue in pharmacovigilance[27], driven by factors such as limited clinical trial sample sizes, homogeneous study populations, short trial durations, and a lack of incentives for healthcare professionals to report adverse events. Furthermore, patients often face dismissal when reporting unrecognized symptoms, leading to delayed identification of new side effects[28]. Historical examples, such as the withdrawal of blockbuster drugs like Avandia (Rosiglitazone, 2010) and Vioxx (Rofecoxib, 2004) due to undetected risks of heart attacks and strokes, highlight the profound complexity of causality assessment and the indispensable need for comprehensive and diverse data sources[29]. Additionally, our framework currently supports only single-drug queries (e.g., "Is headache an adverse effect of metformin?") and does not yet accommodate reverse queries (e.g., "Which drugs cause hand-foot rash?") or class-based queries (e.g., "What oncology drugs cause hand-foot rash?"), which are frequently encountered in clinical practice, especially for drug classes known for specific reactions. For evaluation purposes, we also constrained the LLM output to a binary response for classification comparisons. Finally, the potential for errors in drug name queries due to complex nomenclatures presents another challenge, as documented in studies on existing pharmacovigilance programs.

Future work will aim to address these limitations and expand the framework's capabilities. We plan to integrate real-world, self-reported data from observational databases (e.g., FAERS) and social platforms such as Reddit, X (Twitter), and public comment sections of healthcare-related articles. These sources are invaluable for capturing early signals of both harmful side effects and unexpected therapeutic outcomes often missed by formal reporting systems due to their spontaneous and unfiltered nature. We envision developing a separate conversational assistant interface that leverages near-real-time data for faster detection of unintended effects, thereby supporting both drug safety monitoring and drug repurposing efforts. Augmenting the model with case reports and studies on drug side effects will provide deeper insights into patient-specific factors and relevant literature. For practical use, removing the binary output constraint and utilizing larger reasoning models that can enrich responses with pre-trained drug and side effect information will be crucial. Future iterations will also incorporate semantic search capabilities to recognize mistyped drug names and map generic names to brand names, as SIDER 4.1 uses generic names while patients often query using brand names. Similarly, mapping MedDRA Preferred Terms to synonyms and related medical terminology will ensure patients receive complete information regardless of the terms they use. To mitigate data privacy concerns associated with sharing patient-related information on third-party servers, we will explore deploying our RAG and GraphRAG models on encrypted databases with confidential vector searches on local drives.

**Methods**

Our study aimed to develop and evaluate two Retrieval Augmented Generation (RAG) and GraphRAG frameworks for accurate drug side effect retrieval within Large Language Models (LLMs). The methodologies employed for data acquisition, framework design, and performance evaluation are detailed below.

**Data Source and Preparation**

The Side Effect Resource (SIDER) 4.1 database served as the primary source for drug-side effect associations. This database compiles adverse event reports from randomized controlled trials and post-marketing surveillance. Initial filtering steps were applied to include only drugs with known Anatomical, Therapeutic, and Chemical (ATC) classifications and side effects categorized as MedDRA Preferred Terms (PT). This comprehensive filtering yielded a dataset of 141,209 associations, linking 1,106 marketed drugs to 4,073 unique side effect terms.

To support flexible retrieval mechanisms, the raw SIDER data was processed into three distinct representations (Fig. 1b, Fig. 2a):

- **Data Format A:** This format provides structured, comma-separated descriptions of all known side effects associated with a given drug. An example entry is: "The drug aspirin causes the following side effects or adverse reactions: shock, peptic ulcer, contusion, …".
- **Data Format B:** This format lists each drug-side effect pair on a new line, offering a more granular representation. An example entry is: "The drug aspirin may cause urticaria as an adverse effect, adverse reaction, or side effect".
- **Graph-based Representation:** For the GraphRAG framework, drug-side effect associations were modeled as a graph where drugs and side effects form distinct nodes, and their known relationships are encoded as directed edges. These edges are specifically labeled "may_cause_side_effect," reflecting documented associations. This graph structure was subsequently loaded into a Neo4j graph database.

**Retrieval Augmented Generation (RAG) Framework**

Our RAG system (Fig. 1c, d) was designed to enhance the capabilities of a Llama 3-8B Language Model by integrating it with an external vector database for drug side effect retrieval.

- **Data Indexing:** For text-based retrieval, "Data Format A" was segmented into chunks using a custom algorithm that splits text at newlines. These chunks were embedded into a 1,536-dimensional vector space using the OpenAI ada002 embedding model. This model was selected for its support of up to 8,192 tokens, accommodating the full length of text chunks in "Data Format A," some of which exceed 10,000 characters. The resulting embeddings were then indexed in a Pinecone vector database to enable fast similarity-based retrieval.
- **Query Processing and Retrieval:** When an end-user submits a query (e.g., "Is urticaria an adverse effect of aspirin?"), it is embedded using the ada002 embedding model. This query embedding is then compared against entries in the Pinecone database to retrieve the top five most similar results.
- **Prompt Engineering and Response Generation:** An entity recognition module extracts drug and side effect terms from the user query (e.g., "aspirin" and "headache"). A filtering module then checks whether the extracted drug-side effect pair was present within the top five results retrieved from the vector database. Based on this check, a modified prompt is generated: if a match is found, the prompt specifies that the drug is known to be associated with the side effect; otherwise, it states that the drug is not known to be associated with the side effect. This modified prompt is then passed to a Llama 3 Language Model (8B parameters), which generates a binary YES/NO response indicating the presence or absence of the association. This binary output was specifically chosen because our evaluation framed drug side effect identification as a binary classification problem to predict the presence or absence of an association.
- **Orchestration:** The entire RAG pipeline is orchestrated using AWS Lambda and API Gateway on Amazon Bedrock, ensuring scalability and seamless integration.

**GraphRAG Framework**

Our GraphRAG framework (Fig. 2b) utilizes a graph-based representation for drug-side effect associations to enable precise, relationship-driven queries within a Llama 3-8B Language Model.

- **Graph Database Implementation:** The graph structure, with drugs and side effects as nodes and "may_cause_side_effect" edges, is implemented and managed within a Neo4j database. This database supports efficient querying via Cypher, facilitating rapid traversal and retrieval of relationships.
- **Query Processing and Graph Traversal:** Upon receiving a user query (e.g., "Is headache an adverse effect of metformin?"), an entity recognition module extracts drug and side effect terms (e.g., "metformin" and "headache"). These entities inform the construction of a precise Cypher query (e.g., MATCH (d:Drug {name: 'metformin'})-[r:may_cause_side_effect]->(s:SideEffect {name: 'headache'}) RETURN d, r, s). This query is executed against the Neo4j database to identify whether a directed edge exists between the specified drug and side effect nodes.
- **Prompt Engineering and Response Generation:** The retrieved results from the graph database are processed by a prompt engineering module. If a matching association is found, the prompt is modified to state, "Metformin is known to be associated with headache as a side effect". If no association is found, the prompt states, "Metformin is not known to be associated with headache as a side effect". This prompt modification strategy is identical to that employed in our RAG architecture, ensuring a consistent approach to informing the language model. This refined prompt is then fed into a Llama 3 model (8B parameters), which generates a binary YES/NO response. Similar to the RAG framework, this binary output aligns with our binary classification evaluation approach.
- **Orchestration:** The GraphRAG system is also orchestrated using AWS Lambda and API Gateway on Amazon Bedrock, ensuring scalability and seamless integration for real-time query handling.

**Performance Evaluation**

To rigorously evaluate the effectiveness of our LLM-based system in retrieving drug-side effect associations, we formulated a binary classification task to predict the presence or absence of such associations (YES/NO response). This approach followed prior work on drug side effect prediction[22,23].

- **Evaluation Dataset Creation:** For this evaluation, we filtered the SIDER 4.1 database to include only drugs with at least ten known side effect associations, ensuring sufficient data for robust analysis. For each selected drug, a positive set was constructed by randomly sampling ten known side effects, and a negative set was generated by sampling an equal number of side effects not associated with that drug. This process yielded a balanced dataset of 19,520 drug-side effect pairs, covering 976 marketed drugs and 3,851 unique side effect terms categorized as MedDRA Preferred Terms. This subset was used for assessment due to the computational prohibitive nature of evaluating the full 141,209 associations.
- **Models Assessed:** We assessed the performance of our GraphRAG framework alongside several baseline models:
    - A standalone Llama 3-8B model (without RAG or GraphRAG augmentation).
    - Retrieval-Augmented Generation (RAG) systems utilizing Data Format A.
    - Retrieval-Augmented Generation (RAG) systems utilizing Data Format B.
    - Additionally, to test the generalizability of standalone LLM limitations, we evaluated ChatGPT 3.5 and ChatGPT 4 on a subset of 51 randomly selected drugs.

- **Evaluation Metrics:** Performance was measured across standard binary classification metrics. For a binary classification task, these metrics are defined using the following terms:

  - **True Positives (TP):** Correctly predicted positive instances.
  - **True Negatives (TN):** Correctly predicted negative instances.
  - **False Positives (FP):** Incorrectly predicted positive instances (Type I error).
  - **False Negatives (FN):** Incorrectly predicted negative instances (Type II error).

The formulas for each metric are as follows:

- **Accuracy:** Measures the proportion of correctly classified instances out of the total instances.

$$\text{Accuracy} = \frac{TP+TN}{TP+TN+FP+FN}$$

- **F1 Score:** The harmonic mean of precision and sensitivity, providing a balance between them.

$$F1\ score = 2 \times \frac{Precision \times Sensitivity}{Precision + Sensitivity}$$

- **Precision:** Measures the proportion of true positive predictions among all positive predictions. It answers: "Of all instances predicted as positive, how many are actually positive?"

$$Precision = \frac{TP}{TP + FP}$$

- **Sensitivity (Recall):** Measures the proportion of true positive predictions among all actual positive instances. It answers: "Of all actual positive instances, how many were correctly identified?"

$$Recall = \frac{TP}{TP + FN}$$

- **Specificity:** Measures the proportion of true negative predictions among all actual negative instances. It answers: "Of all actual negative instances, how many were correctly identified?" Specificity=TN+FPTN

$$Specificity = \frac{TN}{TN + FP}$$

# Supplementary Materials

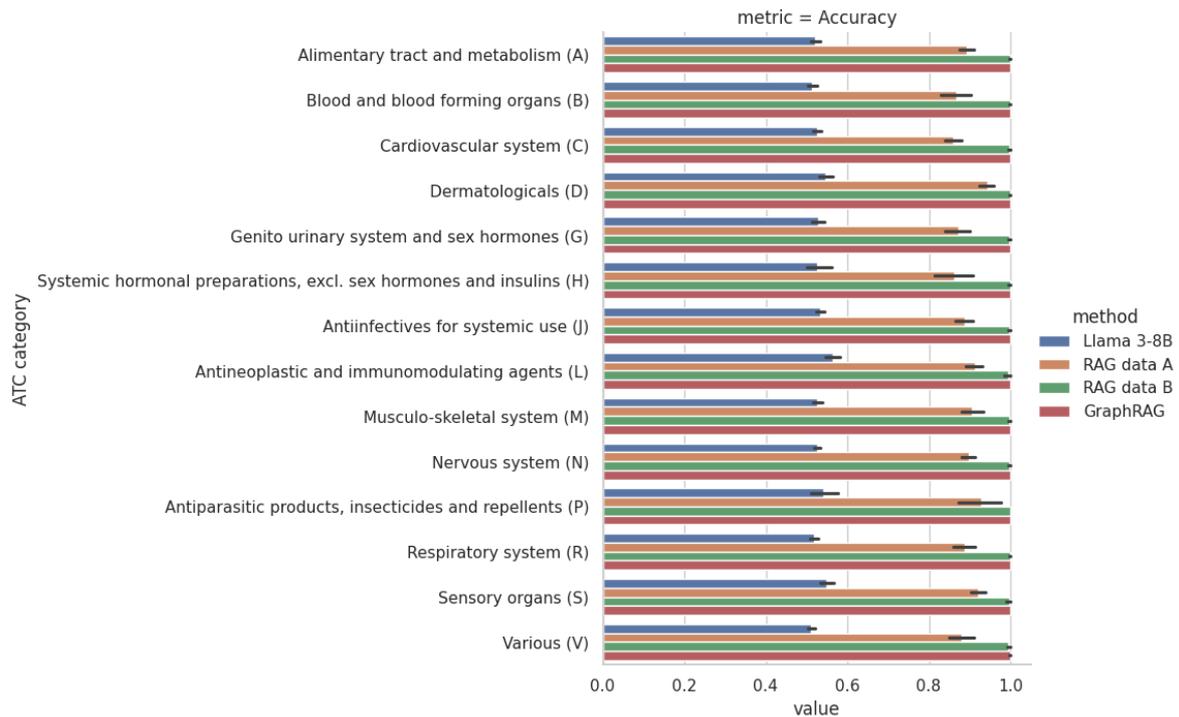

**Supplementary Figure 1. Binary classification accuracy at drug-side effect association retrieval by different methods.** Drugs were grouped by their Anatomical, Therapeutic and Chemical (ATC) class. Methods include standalone Llama 3-8B, RAG data format A, RAG data format B and GraphRAG. The performance was assessed using a balanced dataset of 19,520 drug-side effect pairs, covering 976 marketed drugs and 3,851 side effect terms categorized as MedDRA Preferred Terms obtained from SIDER 4.0.

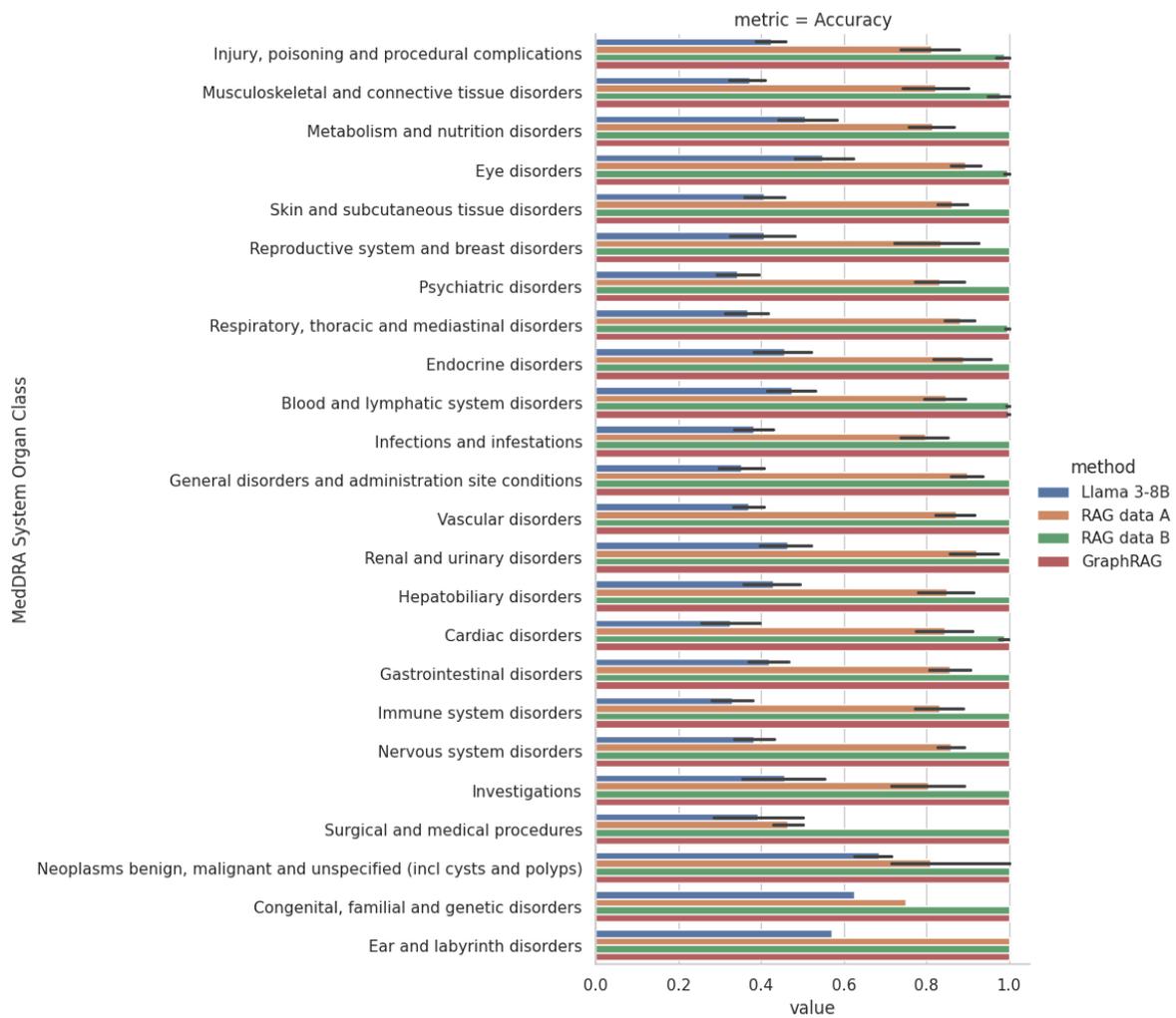

**Supplementary Figure 2. Binary classification accuracy at drug-side effect association retrieval by different methods.** Side effects were grouped by their Organ System class category of disorder according to MedDRA**.** Methods include standalone Llama 3-8B, RAG data format A, RAG data format B and GraphRAG. The performance was assessed using a balanced dataset of 19,520 drug-side effect pairs, covering 976 marketed drugs and 3,851 side effect terms categorized as MedDRA Preferred Terms obtained from SIDER 4.0.